\documentclass[journal=jpcafh,manuscript=article]{achemso}
\usepackage{graphicx}				
\usepackage{amssymb}
\usepackage{amsmath}
\usepackage{physics}
\usepackage{bm}

\SectionNumbersOn
\title{  Schr\"odinger Correspondence  Applied to Crystals}
\author{Eric J. Heller}
\affiliation[Chem]{Department of Chemistry and Chemical Biology, Harvard University, Cambridge, MA 02138}
\alsoaffiliation[Phys]{Department of Physics, Harvard University, Cambridge, MA 02138}
\email{eheller@fas.harvard.edu}
\author{Donghwan Kim}
\affiliation[Chem]{Department of Chemistry and Chemical Biology, Harvard University, Cambridge, MA 02138}

\begin{document}
\begin{abstract}
  In 1926, E. Schr\"odinger published a paper solving his new time dependent wave equation for a displaced ground state in a harmonic oscillator (now called  a coherent state).  He showed that the parameters describing the mean position and mean momentum of the wave packet obey the equations of motion of the  classical oscillator while retaining its width.  This was a qualitatively new kind of correspondence principle, differing from those leading up to quantum mechanics. Schr\"odinger  surely knew that this correspondence would extend to an $N$-dimensional harmonic oscillator. This Schr\"odinger Correspondence Principle is an extremely intuitive and powerful way to approach many aspects of harmonic solids including anharmonic corrections.
\end{abstract}

\section{Introduction}
\begin{figure}[htbp] 
   \centering
   \includegraphics[width=4in]{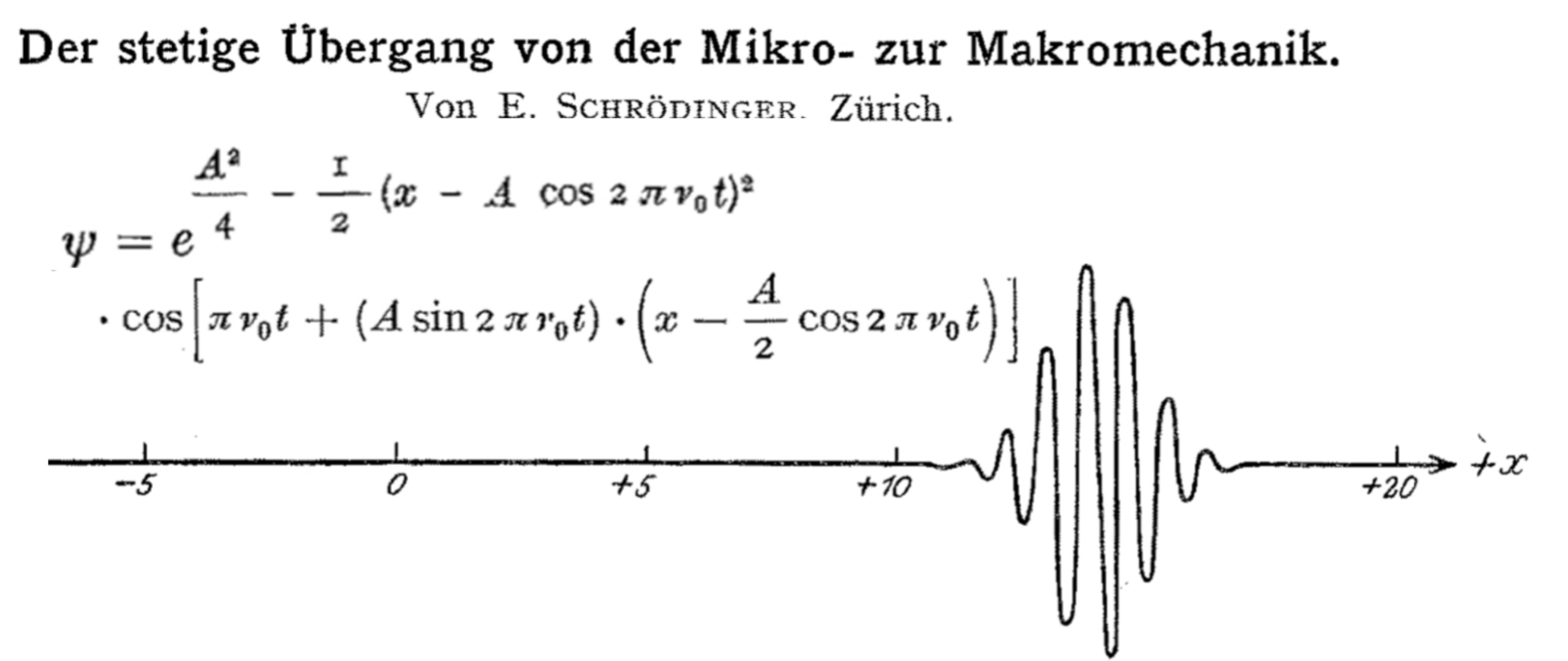} 
   \caption{Photocopy from Schr\"odinger's 1926 paper}
   \label{fig:schr}
\end{figure}
In 1926  Schr\"odinger made the connection between the dynamics of a displaced quantum ground state Gaussian wave packet in a harmonic oscillator  and  classical motion in the same harmonic oscillator\cite{schr} (see figure~\ref{fig:schr}).   The mean position of the Gaussian (its guiding position center) and the mean momentum (its guiding momentum center) follows classical harmonic oscillator equations of motion, while the width of the Gaussian remains stationary if it initially was a displaced (in position or momentum or both) ground state. This classic ``coherent state'' dynamics is now very well known\cite{Heller:way}.  
Specifically, for a harmonic oscillator with Hamiltonian $H =
p^2/2m+ \frac{1}{2} m \omega^2 q^2$
, a Gaussian  wave packet that beginning as 
\begin{equation}
\label{timeG}
\psi(q,0)   =   \exp\left [  i \frac{A_0}{\hbar} (q-q_0)^2 + \frac{i}{\hbar} p_0(q-q_0) + \frac{i}{\hbar} s_0 \right ]
\end{equation}
becomes, under time evolution, 
\begin{equation}
\label{timeG1}
\psi(q,t)   =   \exp\left [  i \frac{A_t}{\hbar} (q-q_t)^2 + \frac{i}{\hbar} p_t(q-q_t) + \frac{i}{\hbar} s_t \right ].
\end{equation}
where
\begin{eqnarray}  \label{harmclass}
p_t &=& p_0\, 
\cos(\omega t) - m\omega q_0\, \sin(\omega t) \nonumber  \\
q_t &=& q_0\, \cos(\omega t) + (p_0/m\omega)\, \sin(\omega t)
\end{eqnarray} 
(i.e. classical motion for the guiding trajectory ($q_t, \, p_t$)) and
if $\, A_0= im\omega/2$ then $A_t=A_0$.
The phase then obeys
\begin{equation}
\label{phseact}
s_t = s_0 +  \frac{1}{2}[p_tq_t-p_0q_0-\omega t].
\end{equation}

More general ``squeezed state'' time dependent solutions apply if the Gaussian spread of the initial  wave packet is a narrower or broader Gaussian than is the ground state of the same oscillator.  In this case, the Gaussian ``breathes,'' becoming wider and narrower twice per period of the harmonic oscillator. Arbitrary time dependent harmonic potentials have exact solutions in terms of Gaussians and their parameters, which are always classical in nature. Here we can exploit the coherent state solutions.

Any imaginable classical motion of a harmonic solid must have a quantum analog, according to   Schr\"odinger correspondence. Each normal  mode of a finite crystal, no matter what the size,  is   a   one dimensional oscillator, with  frequencies and   ``classical'' parameters  appropriate to the initial conditions.  For example, below we will consider  the interesting events after an atom in a crystal is kicked by a $\gamma$ ray photon in the M\"ossbauer effect.

In the case of a general Gaussian in a general, N-dimensional time dependent quadratic form potential, the governing equations are \cite{Heller:way}

\begin{eqnarray}  
\label{gau}
\psi_t(\boldsymbol q)&\equiv& \psi_t(\boldsymbol q_t,\boldsymbol p_t;\boldsymbol q)  = \\ 
&&\exp[{ {\frac{i}{\hbar}}\{  ( \boldsymbol q -  \boldsymbol q _t)\cdot {\boldsymbol A}_t
	\cdot (  \boldsymbol q -  \boldsymbol q _t)
	+  \boldsymbol p _t\cdot (  {\boldsymbol q} -  {\boldsymbol q_t}) +  s_t\}} ]  \nonumber
\end{eqnarray} 
where $ {\boldsymbol A}_t$ is an N$\times$N-dimensional matrix for N coordinates, and 
${\boldsymbol q},\, {\boldsymbol q_t},\, {\boldsymbol p_t}$ are N-dimensional vectors, that 
obey
\begin{eqnarray}  
\label{summary}
{\frac{d}{dt}}\boldsymbol q_t&=& \,\nabla_p H \\
\label{summary2}
{\frac{d}{dt}}\boldsymbol p_t &=& \, -\nabla_q H \\
\label{summary3}
{\boldsymbol A_t} &=& {\frac{1}{2}}{{\boldsymbol P_Z}\cdot {\boldsymbol Z^{-1}}} \\
\label{summary4}
{\frac{d}{dt}}\left(\begin{array}{c}
{\boldsymbol P_Z}\\
{\boldsymbol Z}
\end{array}\right)
&=& \, \left( \begin{array}{cc}
{\boldsymbol 0}&{-{\boldsymbol V''}{(t)}}  \\
{\boldsymbol m^{-1}}&{\boldsymbol 0} \end{array}
\right )\left(
\begin{array}{c}
{\boldsymbol P_Z}\\
{\boldsymbol Z}
\end{array}
\right)\\
\label{summary5}
\dot s_t &=& \,{ L_t} + {\frac{i\hbar}{2}} \textrm{Tr}[ \dot {\boldsymbol Z}\cdot{\boldsymbol Z^{-1}}]
\end{eqnarray} 
or,
\begin{eqnarray}  
\label{stee}
s_t &=&\, s_0 + S_t + {\frac{i\hbar}{2}} \textrm {Tr}[ \ln{\boldsymbol  Z}]
\end{eqnarray} 
${\boldsymbol V''}$ and ${\boldsymbol m^{-1}}$ are N-dimensional matrices of mixed second derivatives 
of the Hamiltonian with respect to position and momentum coordinates, respectively.
That is, 
\begin{eqnarray}  
\left[ {\boldsymbol V''}\right ]_{ij} = {\frac{\partial^2 {  H}}{\partial q_i\partial q_j}}
\end{eqnarray} 
and so forth.
Also
\begin{eqnarray}  
\label{pzee3}
\left(
\begin{array}{c}
{\boldsymbol  P_{\boldsymbol Z_t}}\\ { \boldsymbol Z_t}
\end{array}\right)
&=& \, { \boldsymbol M}{(t) }
\left(
\begin{array}{c}
{ \boldsymbol P_{\boldsymbol Z_0}}\\ 
{ 1}
\end{array}\right)
\end{eqnarray} 
where from equation~(\ref{summary4}) and equation~(\ref{pzee3}) we have  
\begin{equation}
\label{cals3}
{\frac{d{{\boldsymbol M}(t)}}{dt}}  = \, \left( \begin{array}{cc}
{\boldsymbol 0}&{-{\boldsymbol V''}{(t)}}  \\
{\boldsymbol m^{-1}}&{\boldsymbol 0} \end{array}
\right ){\boldsymbol M} (t).
\end{equation}
${\boldsymbol M} (t)$ is the stability matrix,  \index{Stability matrix}
\begin{eqnarray}  
\label{pzee4}
{\boldsymbol M} (t) &=& \left( \begin{array}{cc}
{ {\boldsymbol m}_{11}}&{ {\boldsymbol m}_{12}}   \\
{{\boldsymbol m}_{21}}& {{\boldsymbol m}_{22}} \end{array}\right )
\end{eqnarray} 
where ${\boldsymbol m}_{11} = \partial {\boldsymbol p}_t/\partial  {\boldsymbol p}_0\vert_{ {\boldsymbol q}_0}$, ${\boldsymbol m}_{12} = \partial  {\boldsymbol p}_t/\partial q_0\vert_{ {\boldsymbol p}_0}$, $ {\boldsymbol m}_{21} = \partial  {\boldsymbol q}_t/\partial  {\boldsymbol p}_0\vert_{ {\boldsymbol q}_0}$, and $ {\boldsymbol m}_{22} = \partial  {\boldsymbol q}_t/\partial  {\boldsymbol q}_0\vert_{ {\boldsymbol q}_0}$.

These solutions apply even if different and independent harmonic oscillator degrees of freedom are correlated in the Gaussian, as in the Hamiltonian $H = p_x^2/2m + p_y^2/2m + m \omega_x^2 x^2/2 +m \omega_y^2 y^2/2 $ with 
\begin{align*}
    \psi(x,y) = \exp\  [&-\alpha_x (x-x_t)^2 + i p_{xt} (x-x_t) -\alpha_y (y-y_t)^2 + i p_{yt} (y-y_t) \\
    &+ \mu_t (x-x_t)(y-y_t) + i \phi_t].
\end{align*}
Similar equations follow for  time dependent Gaussians multiplied by Hermite polynomials, i.e. wave packet solutions corresponding to displaced excited harmonic oscillator states.

For our purposes, considering a harmonic crystal in   its ground state with all its separable normal modes (nominally phonons, but see below), and disturbances  of this lattice of various sorts, it will be possible to consider time dependent, independent oscillators for each normal mode degree of freedom. 

Starting with a laboratory coordinate for each atom in a solid, supposed to be free floating as on the space shuttle,  we have $3N-6$ internal, normal coordinates and 6 zero frequency modes that arise from translational and rotational symmetry of the whole. As we will be pointing out in subsequent publications, the zero frequency ``external'' modes are ignored at one's peril, because conceptual problems can arise which can lead to ambiguity and  fundamental  flaws in understanding.

 \section{Phonon momentum and wave vector}
 First we establish the notion of {\it phonon mechanical momentum} $\vec p$ and {\it phonon pseudomomentum} $\hbar \vec k$, with {\it phonon wave vector} $\vec k$.  The pseudomomentum and mechanical momentum are independent, as will be justified in the following.  We will see that phonon mechanical momentum  $\Vec{p}$ belongs to the center of mass translation    of the whole crystal, whereas phonon wave vector $\vec k$ gives a pseudomomentum $\hbar \vec k$ that can be independent of $\Vec{p}$ and is not a mechanical momentum.  
 
 It must be emphasized, although obvious,  that phonons are ``of a piece'' with the lattice. They are parasitic and cannot exist outside the lattice.  In his well known textbook Kittel \cite{kittel}  states ``. . . a phonon really carries zero momentum....'' Kittel was using the word phonon to mean a crystal normal mode, not a corpuscle, just as in quantum electrodynamics, where the word photon is sometimes used for a singly occupied cavity mode.   Ashcroft and Mermin\cite{ashcroft1976solid} clarified the role of corpuscles, stating ``Usually the language of normal modes is replaced by a corpuscular description...'', adding that the concept of a phonon is deliberately analogous to that of a photon. 
 
 The understanding of phonon pseudomomentum vs. mechanical momentum is star-crossed at best. The problem is traceable to the tradition of suppressing the role of whole crystal momentum, usually by  eliminating it before equation 1 so to speak, by writing interactions in internal coordinates. Writing that pseudomomentum is sometimes called ``crystal momentum'', Kittel calls the momentum confusion  a ``delicate point''  in a footnote, stating
 \begin{quote} 
.... Thus in an ${\rm H}_2$ molecule the internuclear vibrational coordinate ${\boldsymbol r}_1-{\boldsymbol r}_2$ is a relative coordinate and does not carry linear momentum; the center of mass coordinate   $({{\boldsymbol r}_1+{\boldsymbol r}_2})/{2}$ corresponds to the uniform mode and can carry linear momentum.
 \end{quote}
 Strangely, Kittel goes on to assign the ``uniform mode''  carrying momentum to the {\it internal} $K=0$ phonon mode, i.e. the infinite wavelength mode of an infinite  crystal. Since he, like almost all authorities, has long since abandoned the crystal center of mass translation, the $K=0$ mode is the best thing left on the shelf so to speak, but it is not correct. This mode  only has the right flavor; the correct assignment is to the $\omega =0$ normal mode, an assignment that is correct for a finite crystal. No internal mode can correctly describe the overall rigid translation of the crystal, which had just been highlighted by Kittel's ${\rm H}_2$ example, which required only 2 atoms!  
 
 Nonetheless Kittel clearly understood the essential momentum carrying role of the center of mass.  He did not assign a wavefunction to this mode, which is actually $\exp[i \vec K \cdot \vec { X}]$, where $\vec { X}$ is the center of mass of the crystal, for which $\vec K\ne 0$ in general.  We call the momentum carrying external modes  ``$K$'' modes, in honor of Prof.  Charles Kittel,  age 102 as of this writing.  We  note that three rotational degrees of freedom are also K modes. It does not matter if the crystal is attached to a lab bench, etc.; momentum conservation is still maintained and the attachment normally has no effect on internal crystal dynamics.
 
 \section{Definitions}
It helps to put some objects in what we call the foreground, and the remainder of the whole crystal is the background.  An electron and/or a phonon may be a foreground object for example. A neutron passing through a crystal is a foreground object.  Foreground objects may have mechanical or pseudo momentum, and so too may the background.  The momentum of the system, i.e. foreground plus background, must be conserved. 

 {\bf Mechanical linear momentum} of an object is defined classically as the sum of masses times Cartesian velocities of the object's constituent particles, and quantum mechanically as the sum over the particle momentum operators $-i \hbar \partial/\partial x_j$. The conservation of this momentum for the totality of objects is a consequence of translational invariance of the total Hamiltonian and its dynamics. Similar statement are made about angular momentum. We will see that phonon mechanical momentum can alternately be assigned to the crystal center of mass momentum or to a localized corpuscle. 
 
 {\bf Phonon pseudomomentum} is  a consequence of Hamiltonian symmetry under   cyclic particle re-labeling. It  need not be conserved if cyclic particle re-labeling is not a symmetry, as when a defect is present.   Phonon pseudomomentum is conserved under phonon-phonon interactions even with anharmonicity. 
 
 {\bf Electron pseudomomentum}  is  a consequence of Hamiltonian symmetry under electron translation by primitive lattice vectors for a fixed lattice.  The pseudomomentum acknowledges that mechanical  momentum  must undulate in energy conserving motion in a periodic potential, replacing  mechanical  momentum which is not conserved with a pseudomomentum that is conserved.
 
 {\bf Crystal pseudomomentum} applies when considering an electron or other particle as a foreground object as distinct from the rest of the crystal, i.e. the background. Then, as the foreground object is held fixed while the background is translated, an undulating potential is again experienced, implying a background crystal pseudomomentum. In elastic defect electron scattering in a crystal, the electron and the crystal exchange pseudomomenta. 
 
 {\bf Total momentum} results from translational symmetry of the whole.  The total momentum $P$ is the sum of the electron pseudomomentum $\hbar k$ and crystal pseudomomentun $\hbar K$: $P = \hbar K + \hbar k$. This is especially apparent if the electron entered from outside the crystal: total mechanical momentum of crystal + electron is conserved, but the total momentum is the sum of two pseudomomenta while the electron is inside the crystal.
 
 \vskip .1in
 \noindent{\it Symmetry and phonon pseudomomentum}
  \vskip .1in
 What symmetry gives rise to phonon pseudomomentum? It is not best thought of as a translational symmetry.  Continuous translational symmetry gives rise to ordinary total momentum and is a symmetry possessed by a finite crystal residing  in free space. The system Hamiltonian is the same under any free translation; it  gives rise to system center of mass momentum conservation. Translation of the system that happens to be  by a lattice translation vector is nothing special; the Hamiltonian is the same even if the translation  is not a lattice vector.    Classical particle shift symmetry is a  more appropriate symmetry:  we replace atoms with their neighbors  in a toroidal geometry, as in $x_n\to x_{n+1}$, with $ x_N\to x_1$.

  For a circular ring of identical atoms with identical forces between them, labeled by their angle $\theta_i$ on the ring, the invariance of Hamiltonian  under the   operator ${\cal R}_+$  can be  written 
  \begin{equation*}
      {\cal R}_+ H(\theta_1,  \theta_2, \dots,\theta_N) = H(\theta_N,  \theta_1, \dots,\theta_{N-1}) {\cal R}_+=H(\theta_1,  \theta_2, \dots,\theta_N) {\cal R}_+ 
  \end{equation*}
i.e. $\theta_i\to\theta_{i-1}$, and ${\cal R}_+ H = H {\cal R}_+$. There is also the inverse operator with the reverse replacement, and a finite cyclic group. This is a permutation symmetry of labels, after which operation the Hamiltonian reads the same.     Cyclic atom replacement requiring ${\cal R}_+ H = H {\cal R}_+$ is certainly not a continuous translation symmetry and it cannot generate a traditional momentum.  The Schr\"odinger equation satisfies
\begin{equation}
{\cal R}_+ H{\cal R}_+^{-1}{\cal R}_+\psi({\boldsymbol \chi})  = H \left [{\cal R}_+\psi({\boldsymbol \chi})\right]= E\left [{\cal R}_+\psi({\boldsymbol \chi})\right ]\end{equation}
with ${\boldsymbol \chi} = (\theta_1,  \theta_2, \dots,\theta_N)$, which implies that eigenfunctions of $H$ are or can be chosen (if they are degenerate) to be eigenfunctions of  ${\cal R}_+$. With the mod $N$ condition, 
\begin{equation}
{\cal R}_+^p\psi({\boldsymbol \chi}) = e^{i p k_m a} \psi({\boldsymbol \chi}).
\label{ruf} 
\end{equation}
where $m$ and $p$ are integers, which implies a pseudomomentum $k_m = 2\pi m/Na$ where $a$ is the distance between neighbors.

  However, this momentum is only defined within multiples of a reciprocal lattice vector, since $k_m \to k_m + 2 \pi M/a$ does not change the  phase factor $e^{i k_m a}$. Thus, in any manipulations we may want to adjust $k$ to put it in the first Brillouin zone by adding multiples of reciprocal lattice vectors.  This is  optional because other Brillouin zones are just copies of the first one. 
 
If some atoms or bonds  differ in some way, ${\cal R} _+ H \ne H {\cal R}_+$ and atom shift symmetry is broken. Phonon wave vector or pseudomomentum will not be conserved. 

The conservation of phonon pseudomomentum for a collection of phonons works like the conservation of mechanical  momentum: the sum of the momenta  must be conserved, although individual momenta may change if the crystal is anharmonic, causing phonons to interact. 
 


%
%
%
 \begin{figure}[htbp] 
   \centering
   \includegraphics[width=4.in]{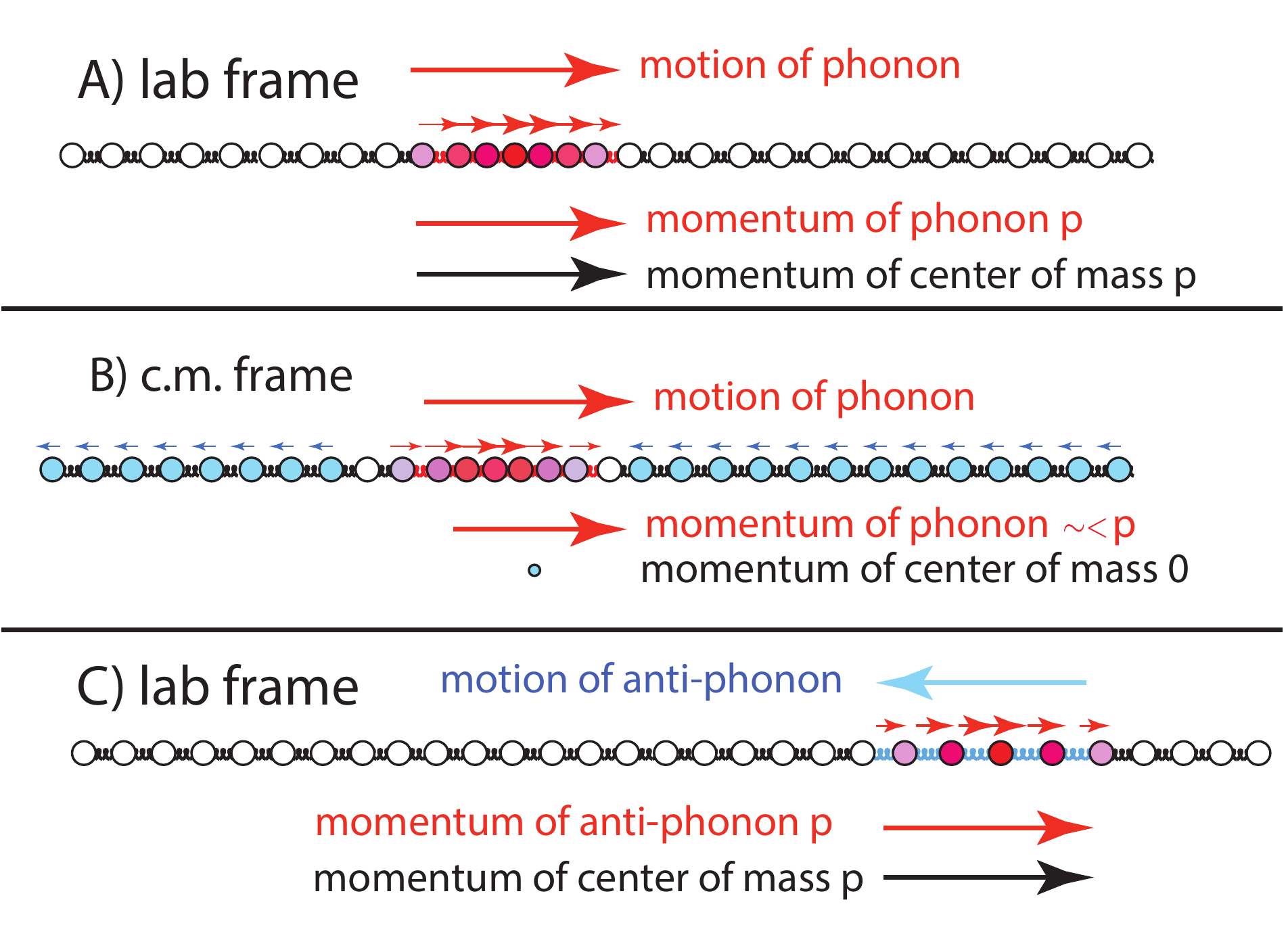} 
   \caption{Three snapshots of a disturbance traveling in a linear chain of equal masses connected by equal force constants.  Snapshots A and B are taken at the same time in different frames.  A is in the lab frame,   B in the center of mass frame at the same time as A, and    C  is taken later, again in the lab frame.  In all frames, springs  that are  stretched relative to their equilibrium length are shown as blue,   and if compressed  are shown as red. Atoms moving right are colored red, moving left atoms are blue with saturation giving speed. The disturbance  in A resulted  from pushing the chain, initially at rest,  from the left end atom.   In the center of mass frame B the motion is composed entirely of  normal modes, none of which has any momentum. A corpuscle (phonon) with locally demonstrable momentum is traveling from left to right, but atoms momentarily not participating in the phonon are drifting in the opposite direction,   so as to make the total center of mass momentum vanish.  Snapshots A  and C are  shown in the lab frame after the same push.      Since atoms are not moving except in the vicinity of the phonon, the center of mass momentum is entirely attributable to the phonon.     In C, we see the situation some moments after the phonon has reflected off the free right hand end. A dilation  has formed and is heading from right to left.  Its momentum is opposite  its direction of motion and its mass density is below the average.  The local momentum contained in the phonon has not changed.  It is    reasonably thought of as an anti-phonon. Following the scenario of  A and C,  on average over time each mass drifts to the right, spending most of its time at rest, only to be rather suddenly displaced to the right as either the negative or positive mass density corpuscle passes by. Over time the chain crawls to the right in an earthworm-like motion. By Schr\"odinger correspondence, we have  just described the quantum mechanical motion as well.}
   \label{fig:corpuscle}
\end{figure}

\section{Line of Atoms}
 Figure~\ref{fig:corpuscle} begins the exercise of Schr\"odinger correspondence to an interesting many body situation. 
 
We now consider an N-atom  1D classical harmonic  chain of atoms (1D crystal), and by Schr\"odinger correspondence a quantum chain.
 
A line or chain of harmonically connected atoms with free ends can be described in the lab or  center of mass frame.  If the  chain is initially at rest in the lab frame and  the  left end atom is transiently forced   left to right,  a corpuscle of energy and momentum is caused to travel down the chain. The system center of mass momentum will have received a boost. In panels A and C of figure~\ref{fig:corpuscle} we describe the  chain  in the lab frame.   Panel B  shows motion in the center of mass frame,  describable entirely in terms of ``internal''  nonon coordinates, none of which carries any momentum. There can be phonon corpuscles with local  momentum density traveling along the chain.  A moment's reflection reveals that any such momentum belongs to the center of mass of the chain; it is often a matter of taste whether the momentum is thought of as belonging to the phonon or to the center of mass.  
 


 In the lab frame used in panels A and C, there is positive center of mass momentum carried by the  K    mode.  All of this momentum  is attributable to the corpuscle.  This remains true after the corpuscle reflects from the end to become an anti-corpuscle or antiphonon of lower than background mass density.  The corpuscle's momentum still points to the right, now in opposition to its motion toward the left.  Both before and after the collision with the free end,  the corpuscle is responsible for the center of mass momentum, which cannot change  in the absence of external forces. When the anticorpuscle collides with the left end, it reflects as a corpuscle again.  The scenario corresponds roughly to an air pressure pulse in a tube with both ends open, apart from the part of the pulse that is radiated from the tube ends.

 The foreground (corpuscle) momentum and the background momentum are different in the two frames.  The corpuscle  in B has a momentum slightly lower than in A and C.  This difference vanishes in the limit of a large chain, since the drift velocity of the non-corpuscle masses  is proportional to the inverse of the chain length.  The time between passes of the corpuscle is proportional to the chain length. The displacement of atoms as the corpuscle passes by is independent of chain length. Over time, the atoms do not drift right or left in the center of mass frame.
 
\begin{figure}[htbp] 
   \centering
   \includegraphics[width=4in]{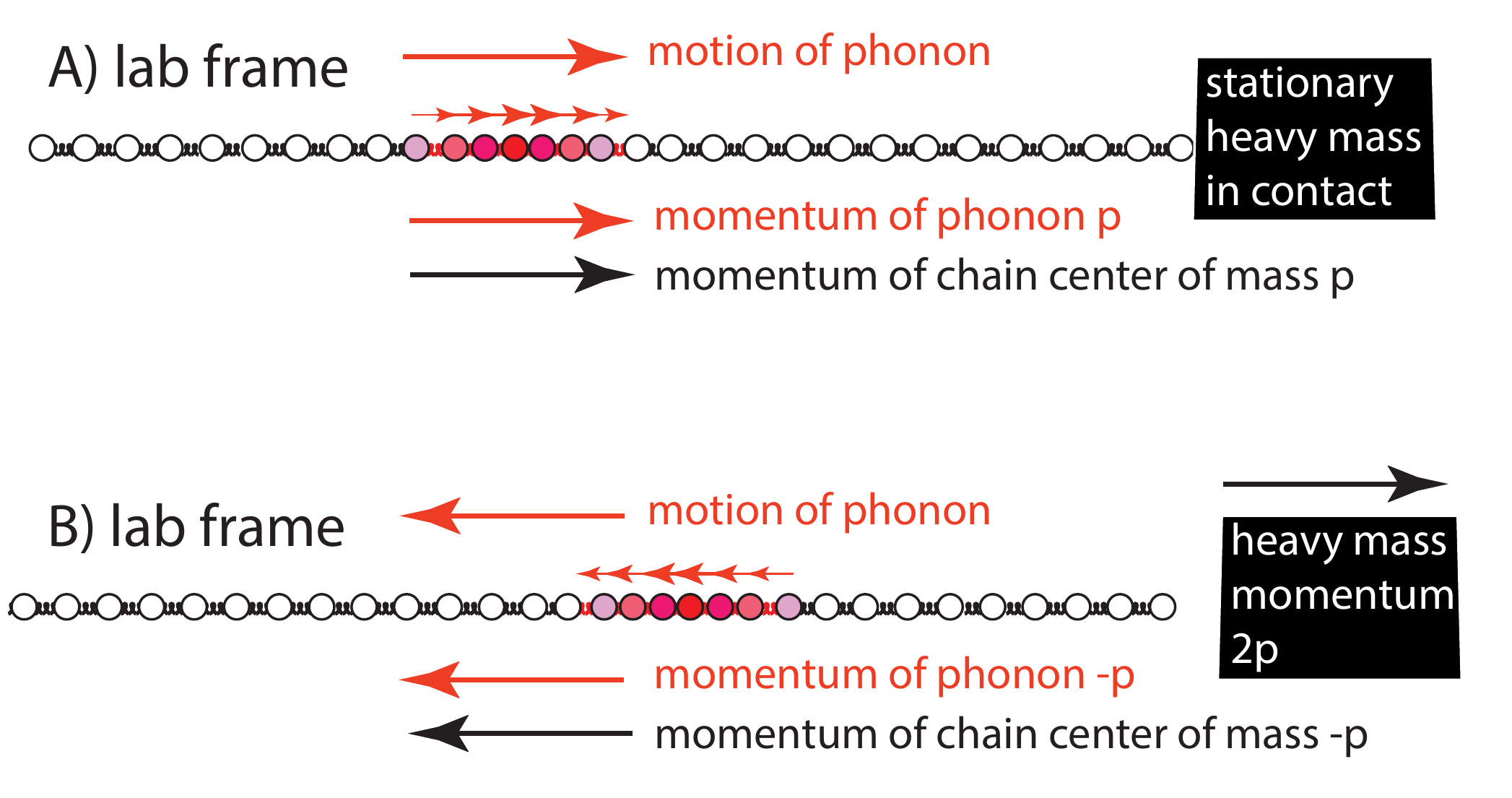} 
   \caption{In A), the initial phonon in figure~\ref{fig:corpuscle} now heads toward  a large stationary mass waiting in contact at the end of the chain. The mass imposes a nearly fixed boundary condition on the right end atom causing reflection of a phonon  in frame B) with reversed momentum in the chain.  The change in chain center of mass mechanical  momentum, $-2 p$, is  balanced by the mechanical momentum gain $+2 p$ of the heavy mass, showing that the phonon momentum was indeed real and of value $p$.  However this phonon has a very ill defined wave vector $k$ and there is no connection here between $k$ and $p$. We emphasize that this classical diagram is every bit quantum mechanical, because of Schr\"odinger correspondence.  }
   \label{fig:phonon}
\end{figure}

 The corpuscle in the lab frame chain has mechanical momentum $p$ that can be calculated in the usual way by adding all the velocities of the masses comprising it.    If a large free mass is just touching the right hand chain member, then very nearly (and exactly as the large mass approaches infinite mass)  momentum $2p$ is imparted to the heavy mass, as the corpuscle arrives(figure~\ref{fig:phonon}). With the heavy mass present, the corpuscle now  recoils  as a positive mass density pulse, with reversed momentum.  The heavy mass behaves  just as if it had been struck by a particle of momentum $p$ and it acquires momentum $2p$.  The total momentum of the chain plus heavy mass remains fixed. 
 

If the chain is pushed to the right   followed quickly by an opposite pull,   a wave of   compression followed by rarefaction is sent down the chain.  The two parts together, if carefully balanced, have cancelling, net zero  momentum, even though they may be traveling briskly down the chain.  Energy  is nonetheless being propagated in this composite phonon.  The pushing and pulling could be done periodically, apart from fading in and out, creating a well defined phonon pseudomomentum for a phonon with no mechanical momentum.
Because of Schr\"odinger correspondence, we can take any classical scenarios quite seriously quantum mechanically. If so, we have learned that phonons can have mechanical momentum or not, quite apart from their pseudomomentum.   We apply the   Schr\"odinger correspondence explicitly to the ring of atoms considered next. 
\begin{figure}[htbp] 
   \centering
   \includegraphics[width=5in]{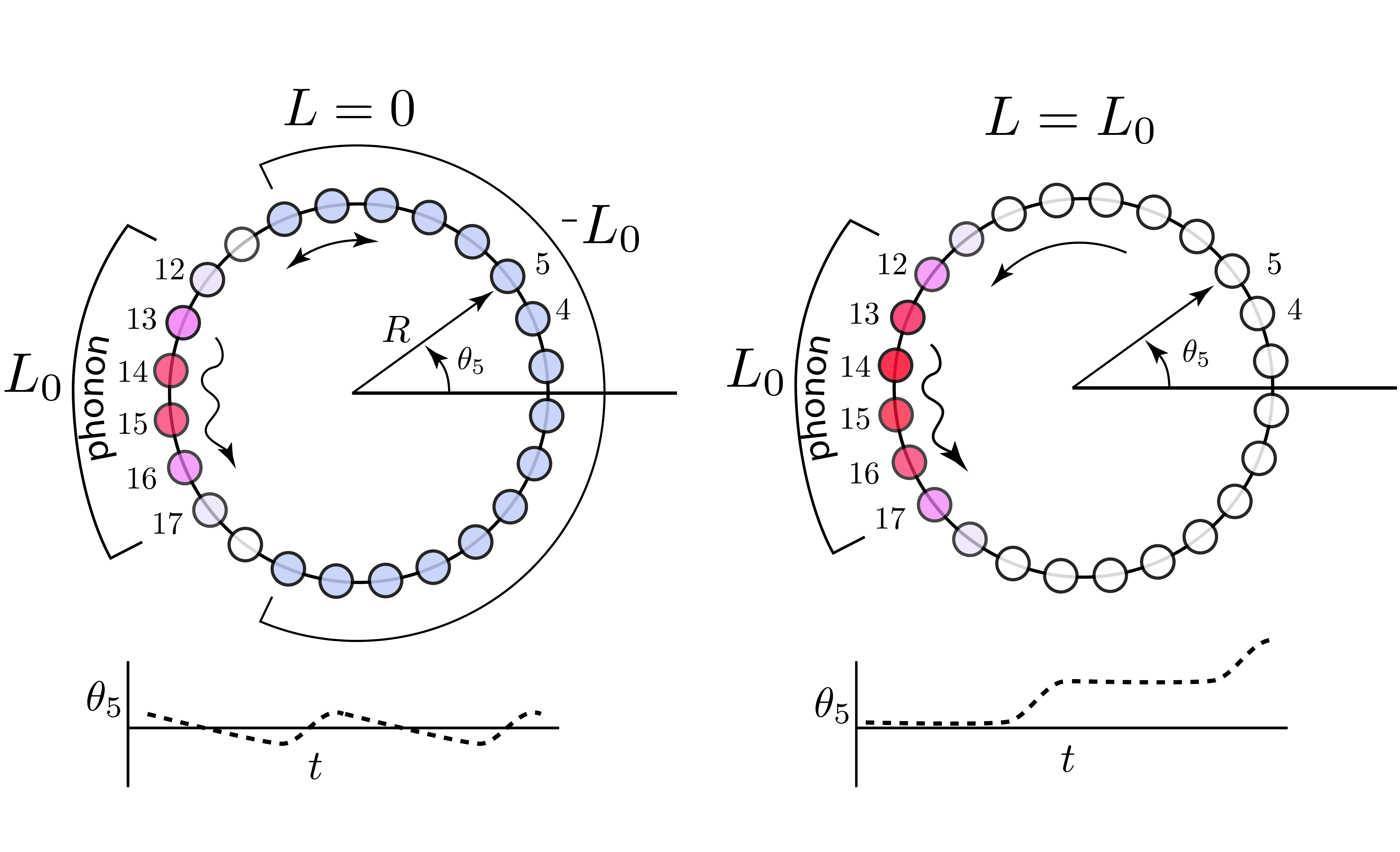} 
   \caption{Atoms with a clockwise drift are shown light blue, counterclockwise is  red.  Steady angle: white.  On the left, the total angular momentum around the center of the ring vanishes, yet a corpuscle travels counterclockwise around the circle.  The corpuscle has angular momentum if considered as a foreground object, with the momentum of atoms 12-17 added together. The background momentum cancels the foreground momentum, giving 0 total angular momentum.  As the corpuscle  passes each atom, the atom suddenly increases its angle counterclockwise; in between such times, the atoms slowly drift clockwise. Over time, there is no drift in angle.  On the right, a component of the zero frequency  K mode (which here is total angular momentum) has been added, just so as to cancel the clockwise drift in between arrivals of the corpuscle. The atoms make rapid counterclockwise  jumps and over time, steady progress counterclockwise in angle. The progress of an atom   with time in each case  is shown at the bottom. On the left, the foreground corpuscle has angular momentum $L_0$, and the background has $-L_0$, total $L=0.$ This is a phonon corpuscle built out of states of 0 angular momentum. On the right, the compensated foreground corpuscle has angular momentum $L_0$, and the background has $L=0$, with total angular momentum $ L_0$. }
   \label{fig:ringcorp}
\end{figure}

\vskip .1in
\section{ Ring of Atoms}
\vskip .1in

A  chain of atoms connected to itself as in a circle (Born-von Karman boundary conditions) is a well known model. Ashcroft and Mermin choose these conditions on an N-atom chain for convenience, stating ``for if N is large, and we are not interested in end effects, then the precise way in which the ends are treated is immaterial.'' \cite{ashcroft1976solid} We  disagree with this seemingly benign statement, for it hides once again  the role of the total momentum or here, the angular momentum. The zero frequency angular momentum mode gets no mention in their treatment.

To be specific, an N atom  chain  slides frictionlessly on a rigid ring confined to the plane.  	Each atom is linked by harmonic springs (figure~\ref{fig:ringcorp}) to its neighbors.  We issue a warning that although this  model has rotational symmetry and well defined angular momentum, it has not been set  free to flex or translate as a whole, so its physics is realistic only for  1D motion along the ring. If for example an atom in the ring were to emit a $\gamma$ ray into the 2D plane, momentum would not be conserved.    

We  form N normal modes  in the usual way.  One of these modes  has zero frequency or no restoring force.  In that   mode, all the atoms move in unison around the ring.  It is the only mode that  can carry angular momentum.

The kinetic energy is that of $N$ atoms each of mass $m$ in a free sliding heavy rigid ring of radius $R$.   There are $N$ angles required to specify the positions of the atoms, which interact by harmonic potentials with a Hamiltonian given as 
\begin{equation}
H=\frac{1}{2 m R^2}\sum\limits_{\ell=1}^N L_\ell^2 +\sum\limits_{\ell=1}^{N-1 } \tfrac{1}{2} \kappa R^2(\theta_{\ell+1}-\theta_\ell)^2,
\end{equation}
which is quantized as 
\begin{equation}
H = -\frac{\hbar^2}{2 m R^2}\sum\limits_{\ell=1}^N \frac{\partial^2}{\partial \theta_\ell^2}+\sum\limits_{\ell=1}^{N-1 } \tfrac{1}{2} \kappa R^2 (\theta_{\ell+1}-\theta_\ell)^2.
\end{equation}


The normal modes $\chi_k$ that diagonalize the force matrix  can be written down by symmetry, for N atoms on a ring of radius $R$,  as 
\begin{equation}
\chi_n = \frac{R}{\sqrt N} \sum\limits_{\ell=1}^{N} e^{i a k_n \ell}\theta_\ell=\frac{R}{\sqrt N} \sum\limits_{\ell=1}^{N} e^{i  2 \pi n   \ell/N  }\theta_\ell,
\label{super}
\end{equation}
\begin{equation}
\theta_\ell = \frac{1}{R \sqrt N} \sum\limits_{n=-b_N}^{b_N} e^{-i a k_n \ell}\chi_n=\frac{1}{R \sqrt N}\sum\limits_{n=-b_N}^{b_N}  e^{-i  2 \pi n  \ell/N  }\chi_n.
\label{super2}
\end{equation}
with $n: -(N-1)/2, \cdots,0,\cdots,(N-1)/2$, $b_N=(N-1)/2$, $a = 2\pi R/N$ and $ k_n = 2n\pi/N a$ with normal frequency $\omega_n = 2\sqrt{\kappa/m}   \sin \left\vert\frac{k_n a}{2}\right \vert =2\sqrt{\kappa/m}   \sin \left\vert\frac{ n \pi }{N}\right \vert $.

We wish to use real valued normal modes $\xi_k$.  The $\chi_0$ mode is already real, so we define $\xi_0 \equiv \chi_0$. We take two real linear combinations of the degenerate modes $\chi_n$ and $\chi_{-n}$ to make two standing waves, one a cosine  $\xi_n=\frac{\chi_n+\chi_{-n}}{\sqrt{2}}=\frac{R}{\sqrt {N/2}} \sum\limits_{\ell=1}^{N} \cos(2 \pi n   \ell/N  )\theta_\ell, n:[ -b_N,-1]$ and one a sine  $\xi_n=\frac{\chi_n-\chi_{-n}}{\sqrt{2}i}=\frac{R}{\sqrt {N/2}} \sum\limits_{\ell=1}^{N} \sin(2 \pi n   \ell/N  )\theta_\ell, n: [1,b_N]$.
The Hamiltonian becomes, in the normal coordinates, 
\begin{equation}
H =  \sum_{n=-b_N}^{b_N}\frac{1}{2 m} p_{\xi_{n}}^2 +   \sum_{n=b_N}^{b_N} \frac{1}{2} m \omega_n^2 \xi_{n}^2
\end{equation}
with normal frequency $\omega_n = 2\sqrt{\kappa/m}   \sin \left\vert\frac{k_n a}{2}\right \vert =2\sqrt{\kappa/m}   \sin \left\vert\frac{ n \pi }{N}\right \vert $.
The coordinate $\xi_0$ stands out as having no restoring force.   It is the K mode, a measure of the overall rotation of the atoms of the ring; its conjugate momentum is proportional to the angular momentum.

 Classically, the total angular momentum reads
\begin{equation}
L = \sum_{\ell=1}^N m R^2\  \dot \theta_\ell = \sqrt{N} m R\ \dot \xi_0 \equiv M R^2 \ \dot\Theta
\end{equation}
and $M=\sum_\ell m$, and $\Theta = \sum_\ell \theta_\ell/N$.
The angular momentum of the ring is the sum of the angular momenta of the atoms.  The angular momentum operator for the ring of $N$  atoms with angle $\theta_\ell$ and of mass $m$ and a ring of radius $R$ is 
\begin{equation}
\hat L = \sum_{\ell=1}^N \hat L_\ell=-i\hbar \sum\limits_{\ell=1}^N \frac{\partial }{\partial \theta_{\ell}} =-i\hbar \frac{\partial }{\partial \Theta} .
\end{equation}

To quantize this system, we take Hamiltonian in normal coordinates (for N odd) 
\begin{equation}
H =  -\sum_{n=b_N}^{b_N}\frac{\hbar^2}{2 m} \frac{\partial^2 }{\partial \xi_{n} ^2 } +  \sum_{n=b_N}^{b_N} \frac{1}{2} m \omega_n^2 \xi_{n} ^2  
\end{equation}
where $$\frac{\hbar^2}{2 m}\frac {\partial^2 }{\partial \xi_0 ^2 }  = \frac{\hbar^2}{2 M R^2} \frac{\partial^2 }{\partial \Theta ^2 }$$

An eigenstate of the chain reads, with angular momentum $\hbar K$ in the (zero frequency) rotational mode, and $\eta_{n}$ quanta in the $n$th  mode, etc.,
\begin{equation}
 \vert {K,  \boldsymbol \eta}\rangle = e^{i K \Theta}\prod\limits_{\substack{n=-b_N\\n\neq0}}^{b_N} \vert \eta_{n}\ \rangle
 \label{st}
\end{equation}
where  
\begin{equation}
\langle \xi_n\vert \eta_{n}\rangle   = \frac{1}{\sqrt{2^{\eta_{n}}\,\eta_{n}!}}   \left(\frac{m\omega_n}{\pi \hbar}\right)^{1/4}  e^{
- \frac{m\omega_n  \xi_{n} ^2}{2 \hbar}}    H_{\eta_{n}}\left(\sqrt{\frac{m\omega_n}{\hbar}} \xi_{n} \right).
\end{equation}

Any $\eta_{n} = 1$   means a singly occupied ``nonon'' mode (we are reserving the term ``phonon'' for corpuscles, as per standard usage). The angular momentum $\hbar K$ with $K$ an integer owes its existence to the K mode free rotation, independent of the  nonon modes, which carry no momentum.  This is obvious if we apply the total angular momentum operator  $\hat L = -i\hbar \partial/\partial {\Theta}$   to the wavefunction of equation~\ref{st}, we get  $\hbar K$, independent of the internal   nonon quantum numbers. Since the function must be the same for $\Theta$ and $\Theta+ 2 \pi $, $K$ must be an integer,  and  the angular momentum is quantized to $L =   \hbar K$. Occupation of any  phonon of coordinate $\xi_{n}$   and wavenumber $k_n, n\ne 0$ contributes no angular momentum, i.e. no mechanical momentum, in spite of superficial appearances (equation~\ref{super}). 

We are not particularly interested in the restriction of allowable angular momenta due to the finite number   N of atoms in the ring.  Rather we take the macroscopic limit of a huge moment of inertia and large N.  Then angular momentum becomes in effect a continuum and we approach the idealization of a large crystal.

There are several ways to make a spatially localized   phonon wave packet.  Perhaps the simplest is to use Schr\"odinger correspondence. Starting with a 0 Kelvin ground state, 

 \begin{equation}
 \vert {0,  \boldsymbol 0}\rangle = e^{i 0 \Theta}\prod\limits_{n=-b_N}^{b_N} \vert 0_{n}\rangle
 \label{st2}
\end{equation}
we apply a  momentum boost to a group of adjacent atoms of the form 
$$ e^{\sum_\ell i p_\ell \theta_\ell/\hbar},$$ imparting angular 
momentum $p_\ell = \hbar K \nu\cdot\exp[- \alpha (\ell-\ell_0)^2]$ to the $\ell^{th}$ atom, creating 
 \begin{equation}
 \ket{\psi_{boost}}= e^{\sum_\ell i p_\ell \theta_\ell/\hbar}\vert {0,  \boldsymbol 0}\rangle = e^{i K \Theta}\sum_{\bm \eta} a_{\boldsymbol \eta} \vert \boldsymbol \eta\rangle 
 \label{st3}
\end{equation}
where $\vert \boldsymbol \eta\rangle $ is shorthand for
$\vert \boldsymbol \eta\rangle =\prod\limits_{\substack{n=-b_N\\n\neq0}}^{b_N} \vert \eta_{n}\rangle.$ The parameter $\nu$ is chosen so that $\sum_\ell p_l=\hbar K$.
The boosted state imparts momentum to each atom, maximizing at the atom $\ell=\ell_0$.
The angular momentum imparted to the center of mass is definite:
\begin{equation*}
    {\hat L}\ket{\psi_{boost}} = \hbar K \ket{\psi_{boost}}.
\end{equation*}

The situation just described corresponds to figure~\ref{fig:ringcorp}, right. A group of atoms has been given a real, mechanical momentum and collectively they have contributed to the total (mechanical) angular momentum. 
\vskip .1in
\noindent{\it  Phonon momentum}
\vskip .1in
In building a localized phonon, any mechanical momentum it possesses accrues to the center of mass momentum and/or angular momentum. However there is a practical sense in which the momentum may be associated with the phonon itself. Indeed we have just created a phonon with mechanical momentum that is attributable to a local disturbance. 

This works by analogy with the standard particle in the box problem, where the sinusoidal eigenfunctions have no momentum, but are a complete set. As such, they can be used in superposition to create a nearly Gaussian wave packet for example with  nonvanishing momentum inside the box. The sleight-of-hand here is that  translational symmetry  is missing in the usual particle in the box problem, and so there is no momentum conservation. Putting the particle in a real box with walls having mass restores translational symmetry. Box + particle states of zero total momentum can be used to construct a particle with considerable {\it local} momentum, with the surrounding  box having canceling momentum. In the case of phonons, we can create states with {\it local} momentum in the lattice using only nonons each with no momentum, but the background or remainder of the lattice will have slow opposite drift, exactly as in figures~\ref{fig:corpuscle} B and ~\ref{fig:ringcorp} A. Indeed by Schr\"odinger correspondence, exactly the same thing happens quantum mechanically.

This is easily arranged by applying a total momentum reverse boost to the momentum change of the atom by atom boost, i.e.
\begin{equation}
 e^{-i K \Theta}\ket{\psi_{boost}}= e^{-i K \Theta} e^{\sum_\ell i p_\ell \theta_\ell/\hbar}\vert {0,  \boldsymbol 0}\rangle= \sum_{\bm \eta} a_{\boldsymbol \eta} \vert \boldsymbol \eta\rangle.
 \label{st4}
\end{equation}
This wavefunction has no total momentum, but atom by atom inspection with the operators
$$\hat p_\ell=-i\hbar \frac{\partial}{\partial \theta_\ell}$$
reveals a group of atoms near $\ell = \ell_0$ with collective net momentum. The sum over all atoms gives zero momentum. This exactly corresponds to figure~\ref{fig:ringcorp}, left. 

The phonon in figure~\ref{fig:ringcorp} does not have a very well defined pseudomomentum, because it is a  pulse with no  modulation. 
\vskip .1in
\noindent{\it  Phonon pseudomomentum}
\vskip .1in
How can a spatially localized  phonon be created, with a well defined  pseudomomentum, near the atom labeled by $\ell_0$?  We can use a linear superposition of one quantum nonons of nearby $k_n$, a narrow range of states of  nearby pseudomomentum.  We will create a state with no total angular momentum (because angular momentum is conserved, and none of the nonons have any) but well defined pseudomomentum. 

  Apart from normalization, we take  
\begin{equation}
\vert \psi_{n,\alpha,\ell_0}({\boldsymbol \xi})\rangle \sim \sum\limits_m  e^{-\tfrac{1}{2 \alpha}(m-n)^2  - i k_{m} a \ell_0} \vert 0 \rangle\vert 0 \rangle\cdots \vert 1_{m}\rangle \vert 0 \rangle\dots\vert 0 \rangle
\end{equation}
or 
\begin{equation}
\psi_{n,\alpha,\ell_0}({\boldsymbol \xi}) \sim \left (\sum\limits_m e^{-\tfrac{1}{2 \alpha}(m-n)^2  - i k_{m} \ell_0} \xi_m \right )e^{-\sum\limits_{n'} m \omega_{n'} \xi_{n'}^2/2\hbar} \equiv \xi_{n,\alpha,\ell_0} e^{-\sum\limits_{n'} m \omega_{n'} \xi_{n'}^2/2\hbar}
\end{equation}
The parameter $k_{n}$ controls the average pseudomomentum of the result; the phonon is centered on the site $\ell_0$. 
The result is   described as:
\begin{equation}
\xi_{n,\alpha,\ell_0}\sim \frac{R}{\sqrt N} \sum\limits_{\ell=1}^{N} e^{i  k_n a   (\ell-\ell_0)  -\tfrac{\alpha}{2} (\ell-\ell_0)^2 }\theta_\ell,
\label{phonon}
\end{equation}
which  leaves atoms with labels far from $\ell_0$   unaltered, but acts like $\xi_n$ near $\ell = \ell_0$.  $\xi_{n,\alpha,\ell_0}$  is a compromise between Fourier extremes of fixed position and fixed wave vector.   This phonon involves only singly occupied nonon modes, by construction.   


 \vskip .1 in

\noindent{\it  Phonon-phonon interaction}
\vskip .1in

Since there is no real, mechanical momentum associated with the pseudomomentum, there is no concern about  momentum conservation  when phonons interact due to anharmonicities or impurities.  Energy must be conserved in any case.  For anharmonicities that are uniform throughout the crystal and intrinsic to the atomic interactions, atom shift symmetry exists and total pseudomomentum  conservation applies. For example, if the energy dispersion is sufficiently linear in the long wavelength acoustic region, two phonons may interact anharmonically to give two new phonons 
$$k_n + k_{n'} \to k_m + k_{m'}$$
where $n+n' = m+m'$. 

 \vskip .1 in
\section{Time evolution of a  M\"ossbauer-like kick}
\vskip .1 in

 	The showpiece of many body elastic quantum transitions is   certainly M\"ossbauer emission, which features a    demonstrably purely elastic component:  no internal  lattice phonon occupations change in a large block of material, despite a sharp kick to one of its atoms.  Only the K-mode is affected in this fraction of elastic events.
																																																	
The sample mass is normally so large that the energy  (due to momentum conservation)  exchanged in a 0-phonon  process is truly infinitesimal.  
A sample of the 	nuclear ground state of $^{57}$Fe placed nearby will hungrily soak up any such $\gamma$-ray with an  enormous resonant cross section.  So narrow is the M\"ossbauer line that moving the nearby  sample at a millimeter per second or less relative to the emitter can Doppler shift the resonance absorption line more than its intrinsic linewidth and let the radiation pass through the sample unhindered, as detected by a counter.  There can be little doubt: some of the emission is totally elastic, it is not just ``quasi-elastic''. 

The remainder of the  $\gamma$-ray emission by the same  nucleus in the same surroundings  gives a broad continuum of   mostly lower energy due to phonon production for   cold samples. Warmer samples  also reveal phonon destruction events. Each emission event, whether totally elastic or not, must be  accompanied by recoil of the sample  with exactly the opposite momentum of the $\gamma$ ray.  There is no escape from whole sample recoil even if phonons are created in the emission. Phonon momentum simply and instantly adds to the sample bulk momentum, as discussed above.

\begin{figure}[htbp] 
   \centering
   \includegraphics[width=2in]{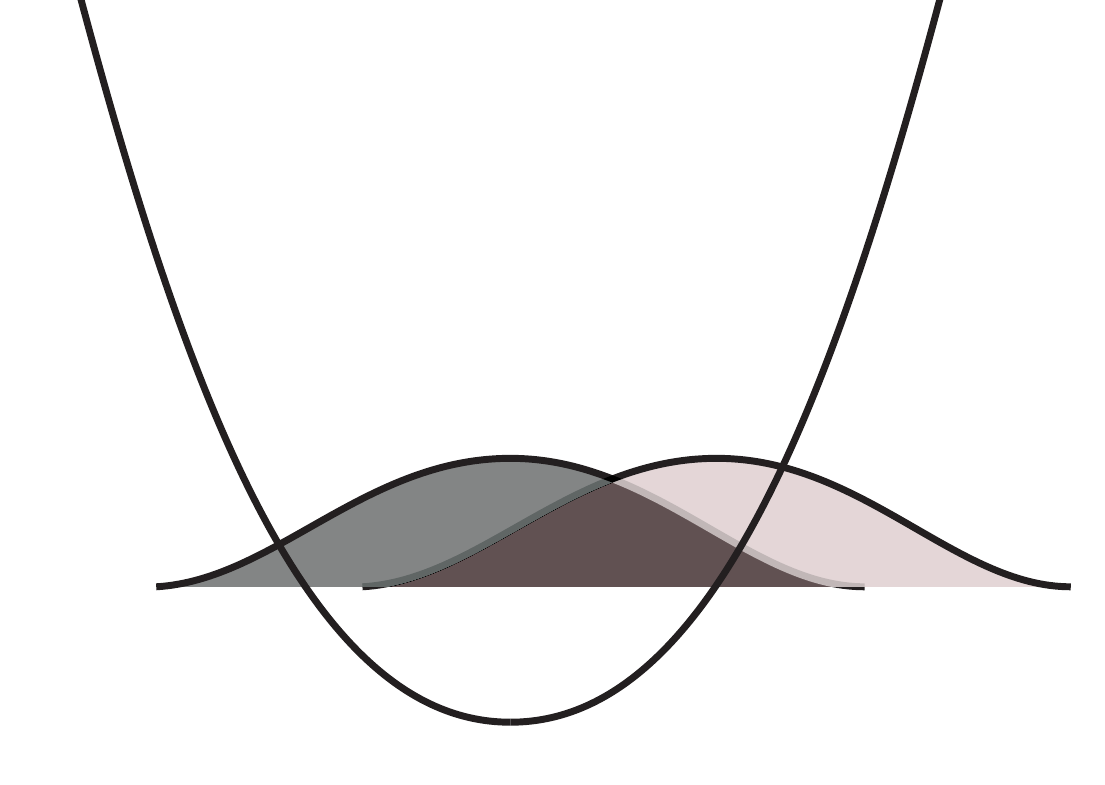} 
   \caption{The ground state Gaussian quantum eigenstate of the harmonic oscillator on the left, centered in the potential, is displaced in position as shown with the lighter colored gaussian displaced to the right. The gaussian has indeed been displaced. Under time evolution, it will oscillate back and forth classically (Schr\"odinger correspondence).  But suppose we ask: what is the probability that the displaced gaussian is still in the undisturbed ground state?  The overlap between the two states shows that this probability is significant, even though classically the function has been displaced with certainty. This is the essence of the M\"ossbauer effect - the wave packet has been boosted in momentum rather than position, just after the $\gamma$ ray emission, but the principle is the same.  What remains is to show that decomposition of this situation into the many body phonon coordinates still leaves a high probability of the totality of all the modes, including the lowest frequency modes, to remain in their ground state.}
   \label{fig:overlap}
\end{figure}

The beginnings (although not a proof) of the quantum effect responsible for the ultranarrow M\"ossbauer line can be seen in a one dimensional oscillator, essentially an Einstein model of a solid (figure~\ref{fig:overlap}). A quantum coherent state wave packet unquestionably oscillates in classical fashion (i.e. Schr\"odinger correspondence) after a kick due to $\gamma$ ray emission.  However if the wave packet energy  is measured, it may be found to be lying quietly in its ground state. The cross section for resonant re-absorption by another nearby   $^{57}$Fe is large, and within about a nanosecond the energy of the $\gamma$ ray (and thus the lattice energy change) can me measured. 

\begin{figure}[htbp] 
   \centering
   \includegraphics[width=5in]{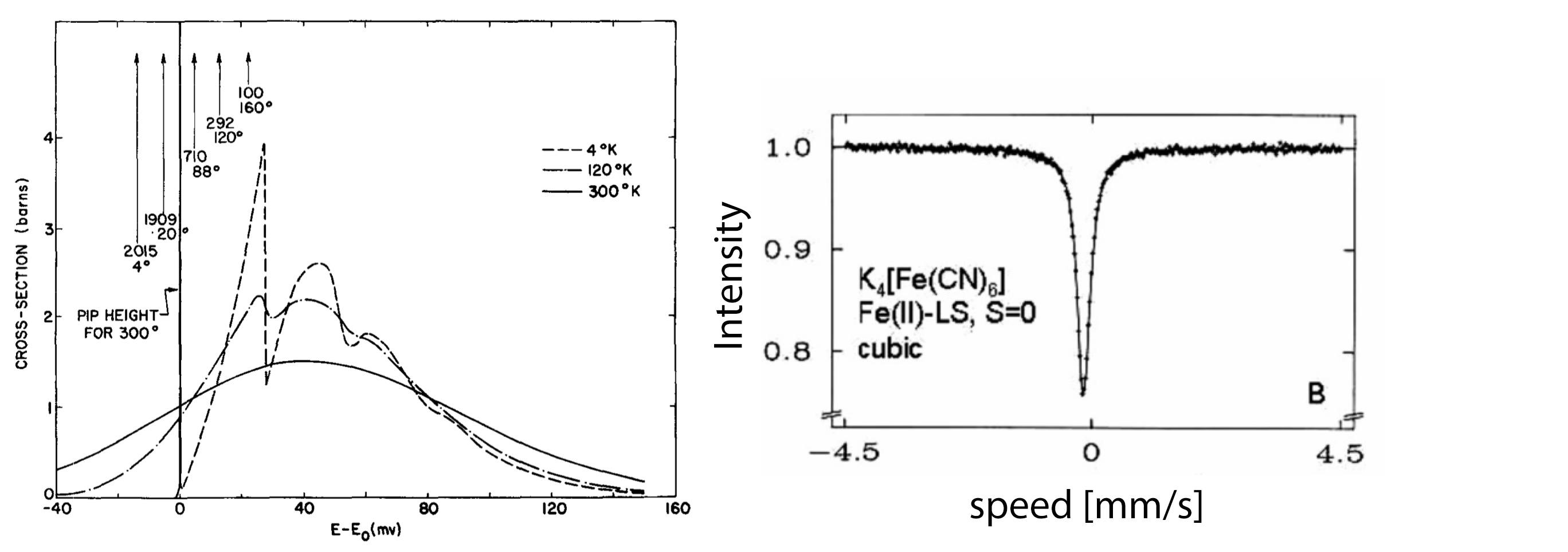} 
   \caption{(left) A figure like this one from 1960 showing the M\"ossbauer line as a ultra narrow ``pip'' on a much broader background is very hard to find under the heading of M\"ossbauer emission, even with a broad Wikipedia search.  The pip height is given as a function of temperature; note that it is about 500 times taller than the ordinate of the graph at 4$^\circ$ K.  It is often the case that ``M\"ossbauer spectroscopy'' is discussed even in  a pedagogical article without beginning with a figure like the one on the left \cite{VISSCHER1960194} showing what is really going on. Almost universally,  M\"ossbauer spectroscopy and even the more fundamental ``M\"ossbauer effect" are shown only   as on the right, spectra that are unremarkable to the eye, unless you notice the x-axis labeled in doppler shifts at speeds of mm/s.}
   \label{fig:moss}
\end{figure}

   The spectrum of energies of the $\gamma$-rays looks like that in figure~\ref{fig:moss}, left.
Discussions of   ``M\"ossbauer spectroscopy" often summarize this beautiful physics only briefly, and rather tend to focus  on something admittedly more practical: the technique of using  tiny shifts of the nuclear resonance energy to probe the chemical environments of the unstable nucleus.  This is  detected  (figure~\ref{fig:moss}) by shifts in the  resonant Doppler   speeds often well under a mm/s, and possible multiple resonance lines. The  shifts reflect   electric  and magnetic field effects  on the nuclear energy level, modulated again by the surroundings of the nucleus and therefore a very useful probe. This is  the business end of M\"ossbauer spectroscopy, and it deserves the attention it gets. But for present purposes, it is the phonon-less recoil that concerns us. This recoil can involve any of the six degrees of freedom of 0 frequency  K modes.


 In addition to the ultra narrow phonon-less M\"ossbauer line, there is a broad continuum of $\gamma$ ray emission that is of lower energy  because a phonon (or more than one) has been deposited in the crystal.   If so,  there are internal agents set loose in the crystal, a phonon  or  phonons. Again, the  whole crystal momentum has jumped by an equal and opposite amount to the $\gamma$ ray.  That amount is slightly lower in magnitude for the inelastic component- the $\gamma$ ray is slightly lower frequency if phonons have been created. 
 
 The wave vector  of the momentum to be deposited  is about $70 $ \AA$^{-1}$.  Sometimes  this is deposited into total crystal recoil without a phonon, i.e. the M\"ossbauser line. If a phonon is created, it need not be responsible for all the recoil momentum.  Indeed,   the momentum can be divided between background  recoil and phonons (if we choose to put them in the foreground), the sum of which   is total system recoil momentum, which is exactly equal and opposite to the photon momentum, now very slightly reduced due to the energy deposited in phonons.  
  
The M\"ossbauer effect continues to surprise.  For example, a clever way of partitioning   the $\gamma$ ray emission into pulses was proposed in reference \cite{Vagizov2015}.

 \begin{figure}[htbp] 
   \centering
   \includegraphics[width=3in]{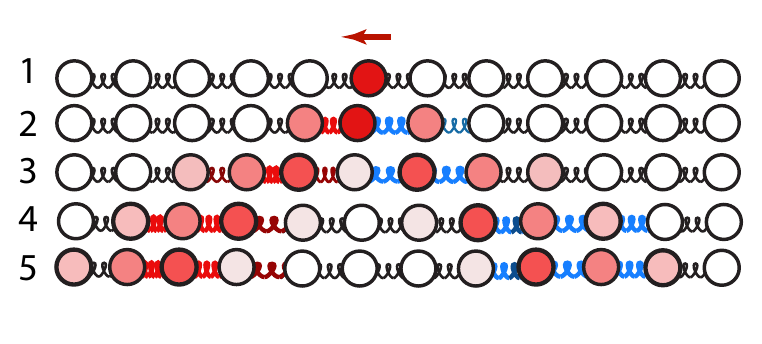} 
   \caption{ An impulsive kick to one atom (as in M\"ossbauer $\gamma $ ray emission) in a harmonic chain produces a phonon-antiphonon pair, seen here propagating in five snapshots at increasing times. The motion is indicated by red atoms if they are traveling right to left, by red springs if the springs are compressed relative to equilibrium,  and by blue springs if they are extended. At the top, the $m^{th}$ atom has just been accelerated (kicked) but there is no displacement as yet. In step 2, the atom to its left has been accelerated and moved slightly, and the atom to its right has been pulled to the left. This sets up a cascading propagation of disturbances in both directions. A dilation can be seen heading to the right, and a compression to the left. Both the phonon and the antiphonon have right-to-left momentum with atoms   moving to the   left as either disturbance passes by.  In this scenario, the atoms become permanently displaced. By the Sch\"odinger correspondence, this scenario has direct quantum counterparts. This figure omits high local back and forth oscillation with no momentum transfer consequences; see figure~\ref{fig:moden}.}
   \label{fig:pair}
\end{figure}

The proper derivation of the M\"ossbauer $\gamma$ ray spectrum is given by Maradudin\cite{maradudin}. Here we will examine an ersatz M\"ossbauer kick, by providing an instantaneous momentum boost to a single atom.  The exact momentum supplied to the system is therefore given, whereas it varies somewhat in  $\gamma$ ray emission according to whether or not a phonon or phonons  accompany the emission.

Suppose $ \psi_{n,\alpha,\theta_0}({\boldsymbol \xi}) $ is a lattice that just got a momentum kick $\hbar k$ to an atom at position $\ell_m$, what happens under time evolution?
We take phonons  to be absent initially, and suppose no initial center of mass momentum. The initial many body wavefunction is then (apart from normalization)
\begin{equation}
\psi({\boldsymbol \xi},0)  = e^{-\sum_{n} m \omega_{n} \xi_{n'}^2/2\hbar} 
\end{equation}
Each   ground state   nonon mode simply evolves with a phase factor appropriate to its energy, but all these phase factors add in the exponent and nothing but an overall  phase develops.  This is another way to say that the ground state is stationary.
After the $\ell^{th}$ atom gets its kick at $t=0$, it becomes
\begin{equation}
\label{kicked}
\psi_{\ell k}({\boldsymbol \xi},0)  = e^{-\sum _{n} m \omega_{n} \xi_{n}^2/2\hbar } \ e^{i k R \theta_{\ell}} 
\end{equation}
We can ask the following questions now and as time evolves:

1) What is the momentum of the center of mass?

2) What is the time evolution of the many body wavefunction?  

3) Where is the momentum located on the chain?

4) What is the probability that the kick produced no phonons? (i.e. was elastic)

5) What is the distribution of other phonon probabilities?

Only the answers to 2) and 3) will depend on time if the system is not further disturbed. We now address these questions in turn.

(1) The momentum delivered to one atom was~$\hbar k$.  Since there was no center of mass momentum to begin with, we had better find the momentum of the center of mass is now also $\hbar k$.  The atom with coordinate $\theta_{\ell_0}$ is expanded as 
\begin{equation}
\theta_{\ell} = \sum\limits_{n=0}^{N-1} U_{\ell n} \xi_n = \frac{1}{R\sqrt N} \xi_0 + \sum\limits_{n=1}^{N-1} U_{\ell n} \xi_n
\end{equation}
where $ \xi_0 = \frac{R}{\sqrt N}\sum\limits_{j=1}^{N} \theta_{j}.$
The center of mass coordinate for $N$ identical masses along a line at positions $\theta_{j}
$ is 
\begin{equation}
\Theta = \frac{1}{N} \sum\limits_{j=1}^{N} \theta_{j} = \frac{1}{R\sqrt N} \xi_0.
\end{equation}
The wavefunction equation  \ref{kicked} at $t=0$ with the kick $ e^{i k R\theta_{\ell}} $ has the term $e^{i k \frac{1}{\sqrt N} \xi_0} = e^{i k \Theta}$ proving that the center of mass acquired the correct momentum. 

(2) To find the time evolution of the many body system, we use  Schr\"odinger correspondence and the equations
\begin{eqnarray}
\xi_{n t} =\frac{\hbar k_n}{m \omega_n} \sin{\omega_n t} \nonumber \\
p_{nt} = \hbar k_n\cos(\omega_n t)
\end{eqnarray}
where $k_n = k\  U_{\ell n}$ and the wavefunction evolves as 
\begin{equation}
\psi_{\ell k}({\boldsymbol \xi},t)  = e^{-\sum _{n} m \omega_{n} (\xi_{n}-\xi_{n t})^2/2\hbar  + i p_{ n t}(\xi_{n}-\xi_{nt})/\hbar + i \phi_t}. 
\label{evolve}
\end{equation}
 We won't bother to specify the overall phase here. 
 
(3) To ask where in the chain does the momentum density lie, we need to examine the atomic momentum operators
\begin{equation}
\hat p_{j} = -i\hbar \frac{1}{R} \frac{\partial}{\partial \theta_j}.
\end{equation}
 At $t=0$, equation  \ref{kicked} reveals of course that the (real) momentum  resides locally on the $\ell^{th}$ atom. We   want to discuss the situation as time evolves.

\begin{figure}[htbp] 
    \centering
    \includegraphics[width=5in]{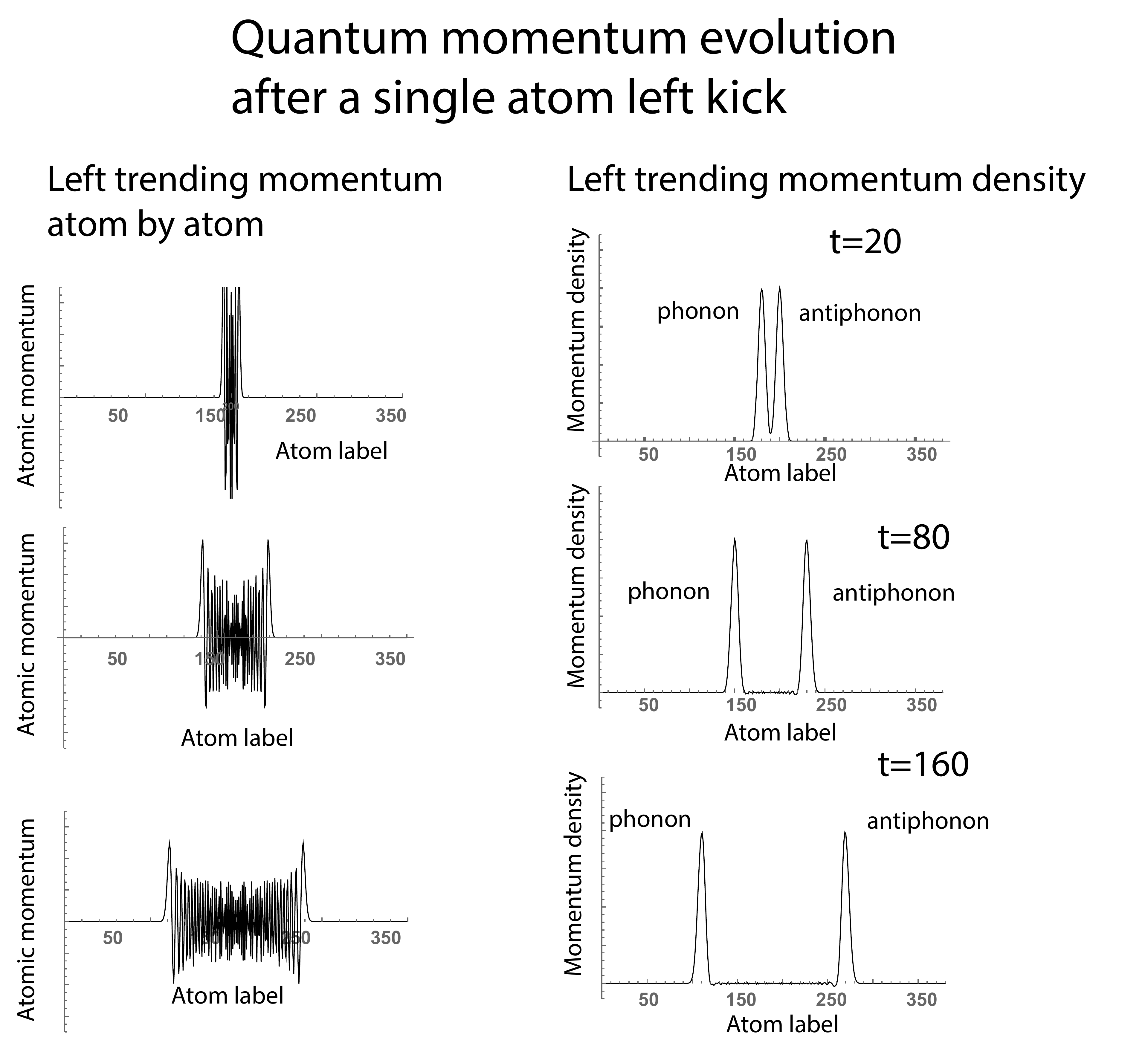} 
    \caption{Here the central atom was kicked to the left and the positive values in the plots are left heading momenta.}
    \label{fig:moden}
 \end{figure}

 We  apply the momentum operator for each atom separately and plot the results atom-by-atom (figure~\ref{fig:moden}, left column) or averaged over a range of neighboring atoms (figure~\ref{fig:moden}, right column) to get a momentum density.  Here the central atom was kicked to the left and the positive values are left heading momenta.  Note that the  phonon, anti-phonon narrative is confirmed. This is not a semiclassical result of some sort; by Schr\"odinger correspondence, it is the exact result.
 
The group velocity of the phonon is $v_g\approx \delta \omega/\delta k \approx a \sqrt{\kappa/m}$ at long wavelengths.

Schr\"odinger correspondence and the classical mechanics of the chain informs us  that the kick produces a phonon-antiphonon pair   (figure~\ref{fig:pair}). The phonon and the antiphonon travel in opposite directions, departing the site of the kick, but carrying the same momentum.  This is certainly different than the standard picture of a ``phonon'' heading off in one direction, carrying all the momentum.  The classical reason for the pair production is clear from figure~\ref{fig:pair}, and by Schr\"odinger correspondence it is clear that the classical picture is extremely relevant to the quantum evolution.

  (4)  What is the probability that the kick produced no phonons? This is given by the square amplitude, in the case of an initial quiescent ring, as 
\begin{eqnarray}
P_{elastic}&=&\vert \langle \psi_0({\boldsymbol \xi})\vert e^{i k R \theta_\ell}\vert \psi_0({\boldsymbol \xi})\rangle\vert^2 \nonumber \\
  &=& \left \vert R^N \int d \theta_\ell \   e^{i k R \theta_\ell}\cdot\int d \theta_1\dots d \theta_{\ell-1} d \theta_{\ell+1}\dots   \langle \psi_0({\boldsymbol \xi})\vert \psi_0({\boldsymbol \xi})\rangle \right\vert^2 \nonumber \\
   &=& \left \vert R\int d \theta_\ell \  \sqrt{\frac {w}{\pi}}e^{i k R \theta_\ell }e^{-w R^2\theta_\ell^2}\right\vert^2\nonumber \\
   &=& e^{-k^2/2w} \equiv e^{-2 W}.
  \end{eqnarray}
The elastic fraction $e^{-2 W}$ is called a Debye Waller factor and is clearly related to the root mean square dispersion of the impacted atom. This result carries over to finite temperature. The finite elastic fraction for M\"ossbauer emission in a 3D solid at finite temperature  is affirmed by the fact that sub angstrom resolution of atoms in STM scans of surfaces is possible.  It would not be if the root mean squared position dispersion of an atom were large. 

(5) The elastic zero phonon line is flanked by a density of inelastic processes that have an energy sprectrum given by the Fourier transform of the autocorrelation of the kicked state:
\begin{eqnarray}
\Sigma(\omega) = \int dt \ e^{i \omega t} \langle\phi\vert \phi(t)\rangle
  \end{eqnarray}
  where $\vert \phi\rangle =  e^{i k R\theta_\ell}\vert \psi_0({\boldsymbol \xi})\rangle$ and $\vert \phi(t)\rangle =  e^{-i H t/\hbar }\vert \phi\rangle.$ The autocorrelation $\langle\phi\vert \phi(t)\rangle$ is evaluated using equation~\ref{evolve} and straightforward Gaussian integrals, but its Fourier transform is problematic  and it  is best evaluated numerically.
 \vskip .1 in
 
\section{ Indirect transitions in solids}
\vskip .1 in

We have seen   that phonon wave vector and phonon pseudomomentum can be  unrelated.  Of course a dispersion relation holds between phonon wave vector (and pseudomomentum) and phonon energy, i.e. $E = \hbar \omega(\vec k)$. However in some important situations, such as indirect interband transitions in solids, phonon pseudomomentum has to cancel electron pseudomomentum changes in order to keep pseudomomentum conserved.  Electron pseudomomentum  (pseudo because of the periodic lattice potential) is  closely related to mechanical momentum.  An electron has a life outside of the crystal and with enough pseudomomentum it can escape the crystal with definite mechanical momentum.  The renormalization of  electron momentum inside a crystal is akin to the changes of refractive index and light speed in a crystal. 

Indirect transitions require that an electron pseudomomentum change be matched by opposite  pseudomomentum associated with a phonon.   How does this happen, if the connections between phonon wave vector and momentum are so loose?

Raman scattering in polyacetylene is one of the clearest  examples,  with a solid experimental foundation dating back at least to the 1980's. However the   unusual Raman spectrum of polyacetylene was not understood until recently.  An indirect transition was found  responsible for the mysterious side band that shifts with incident laser frequency\cite{poly}. Starting in the equilibrium ground state, a UV photon promotes a valence band electron resonantly to the conduction band.  Normally, given the $k\approx 0$ nature of the photon, the electron would share the same conduction band and valence band quasimomenta. In this elastic scenario, no phonons would be involved.  In a minority of cases the promotion is accompanied by the production or destruction of a phonon.  This is known because when a photon is finally emitted it can be red (Stokes) or blue (anti-Stokes) shifted by a phonon energy, as is fundamental to Raman scattering.  The phonon needs either to be at $k=0$ in an optical mode (in which case the Bloch electron does not change momentum)  or to have twice the wave vector of the electron, which has its momentum reversed in going from valence to conduction bands.     The exactly reverse momentum of the electron preserves its energy,  assisting an otherwise difficult energy-momentum matching exchange. The final step, elastic backscattering of the electron by an edge, defect,  or impurity, makes the electron able  to emit a Raman photon to the hole.  The wave vector of the electron was initially fixed  by the energy and momentum matching requirements of the photoabsorption, including the instantaneous production of a phonon, courtesy of the coordinate dependence of the  electronic transition moment. This is also the scenario leading to the Raman D line in graphene, which indeed requires impurities to perform the last elastic backscattering step and thus ``light up'' the D line\cite{ramanacsnano}.

 The energy-momentum requirements on this resonant absorption event are strict.  However the electronic transition moment has a component that is constant and another  that oscillates at wave vector $2 \vec k$\cite{poly}, facilitating the $k \to -k$  electron quasimomentum reversal and birth of the $2 k$ phonon. The whole process takes place  resonantly; the electronic energy change is slightly reduced from the photon energy to pay for the phonon energy. 
 
 Just as a M\"ossbauer event can transfer momentum to the crystal elastically, it seems plausible that the electron could have its momentum reversed elastically, with the background lattice providing the momentum compensation. The background takes the momentum in any case, and any such elastically born electron, if backscattered, would emit a photon hidden in the Rayleigh emission. If this is happening, it would be buried in the elastic Rayleigh emission and become very difficult to detect.
 
\begin{acknowledgement}

EJH thanks W.P. Reinhardt for many consultations going back 50 years, including his first exposure to the ring of atoms problem, which appeared in his thesis under Prof. Reinhardt\cite{hellerthesis}.  The authors thank Prof. E. Kaxiras for valuable consultations regarding this research. We also thank Siddharth Venkat and Vaibhav Mohanty for informative discussions. 
 .

\end{acknowledgement}
%
\bibliography{elastic}

\end{document}